\documentclass[aps,10pt,twocolumn]{revtex4}
\usepackage{amssymb}
\usepackage{graphicx,setspace}
\usepackage{amsmath}

\begin{document}
\title{Galilean Covariance versus Gauge Invariance}
\author{Germain Rousseaux}
\affiliation{Universit\'{e} de Nice-Sophia Antipolis,
Laboratoire J.-A. Dieudonn\'{e}, UMR CNRS-UNS 6621,
Parc Valrose,
06108 Nice Cedex 02, France, European Union.}

\date{\today}

\begin{abstract}
We demonstrate for the first time and unexpectedly that the Principle of Relativity dictates the choice of the "gauge conditions" in the canonical example of a Gauge Theory namely Classical Electromagnetism. All the known "gauge conditions" of the literature are interpreted physically as electromagnetic continuity equations hence the "gauge fields". The existence of a Galilean Electromagnetism with TWO dual limits ("electric" and "magnetic") is the crux of the problem \cite{LBLL}. A phase-space with the domains of validity of the various "gauge conditions" is provided and is shown to depend on three characteristic times : the magnetic diffusion time, the charge relaxation time and the transit time of electromagnetic waves in a continuous medium \cite{Melcher}.
\end{abstract}

\maketitle

The Standard Model of Physics is based on the assumed existence of a superior principle called Gauge Symmetry which would rule all the laws of Physics: {\it Physical theories of fundamental significance tend to be gauge theories. These  are theories in which the physical system being dealt with is described by more variables than there are physically independent degree of freedom. The physically meaningful degrees of freedom then reemerge as being those invariant under a transformation connecting the variables (gauge transformation). Thus, one introduces extra variables to make the description more transparent and brings in at the same time a gauge symmetry to extract the physically relevant content. It is a remarkable occurrence that the road to progress has invariably been towards enlarging the number of variables and introducing a more powerful symmetry rather than conversely aiming at reducing the number of variables and eliminating the symmetry} \cite{Henneaux}.  Wolfgang Pauli was used to ask at the end of tiresome seminars he attended loosely if the principal result presented by the speaker was "gauge invariant" \cite{Enz}.  Hence, the concept of Gauge Theory has emerged progressively in Physics such that the equations feature variables ("gauge fields") which are underdetermined and in order to remove this degree of liberty ("gauge transformations") a closure assumption ("gauge condition") is formulated \cite{Okun}. Similarly, the Principle of Relativity is known to be a robust safeguard when scaffolding a new theory since the proposed new laws must be covariant with respect to the transformations of space-time. 

The goal of this paper is to remove the Gauge symmetry in the most famous example of a supposed Gauge Theory namely Classical Electromagnetism by revealing a conflict with another symmetry that is the Principle of Relativity. To do so, we first emphasize the Riemann-Lorenz approach to Electromagnetism. Therein the central role is played by the vector and scalar potentials ${\bf{A}}$ and $V$, unlike the Heaviside-Hertz approach, which rather relies on the fields ${\bf{B}}$ and ${\bf{E}}$ themselves (for a justification, see \cite{riemannlorenz} and \cite{A}). In this formulation, the fields are defined as a function of the potentials (and not the reverse) according to ${\bf{B}}=\nabla \times {\bf{A}}$ and ${\bf{E}}=-\nabla V-\frac{\partial {\bf{A}}}{\partial t}$. As a consequence of these definitions and using obvious vectorial identities, the fields obey the following constraints $\nabla .{\bf{B}}=0$ and $\frac{\partial {\bf{B}}}{\partial t}=-\nabla \times {\bf{E}}$. But how are defined the potentials themselves ? They are the mathematical solutions of the Maxwell-Minkowski equations written for the excitations:
\begin{equation}
\nabla .{\bf{D}}=\rho \quad and \quad \nabla \times {\bf{H}}=\frac{\partial {\bf{D}}}{\partial t}+{\bf{J}}.
\end{equation}
We have to relate the excitations to the fields thanks to the constitutive relations for media at rest and then the fields to the potentials thanks to their definitions above. The current density features two terms ${\bf{J}}={\bf{J}}_{constitutive}+{\bf{J}}_{external}$. The constitutive current which expresses the matter response to the fields depends on the medium. For example, in Ohmic conductors, we have ${\bf{J}}_{Ohm}=\sigma {\bf{E}}=\sigma \left(-\nabla V-\frac{\partial {\bf{A}}}{\partial t}\right)$ whereas in a Superconductor \cite{Thinkham}, the constitutive relation becomes ${\bf{J}}_{Supra}=\frac{\hbar ne^*}{m}\left(\nabla \phi - \frac{e^*}{\hbar}{\bf{A}}\right)$. For continuous media at rest the excitations are related to the fields according to ${\bf{D}}=\epsilon {\bf{E}}$ and ${\bf{B}}=\mu {\bf{H}}$. We get a system of equations where the unknowns are the potentials ${\bf{S}} \left( {\bf{A}}, V; \epsilon, \mu, \rho, {\bf{J}} \right)=0$ provided the sources are given or expressed in function of the potentials which vanish far from the latter or take prescribed values on given boundaries. However, the system ${\bf{S}}=0$ cannot be solved unless another equation is added. This closure assumption is usually known as the "gauge condition" in the Heaviside-Hertz formulation since the potentials are {\it de facto} underdetermined (by the "gauge transformations" ${\bf{A}}'={\bf{A}}+\nabla f$ and $V'=V-\frac{\partial f}{\partial t}$ \cite{Okun}) if and only if they are defined in function of the fields and not the reverse as in the Riemann-Lorenz formulation.

In the following, we will show that the closure assumption is a consequence of the Relativistic or Galilean nature of the  problem under study. For that purpose, we will recall the Stratton "gauge condition" which is, according to us, the most general physical constraint which can be used all the times. Then, thanks to the Galilean limits of Classical Electromagnetism \cite{LBLL, Montigny, EPL05, AJP, EPL08}, we will approximate the Stratton "gauge condition" depending on the context and we will recover the other "gauge conditions" introduced in the literature by pointing out their domain of validity.

The Stratton "gauge condition" was introduced in Physics at M.I.T. in 1941 by Julius Adams Stratton \cite{Stratton} to cope with the propagation of electromagnetic waves in Ohmic conductors such that the sources are given by $\rho =0$ and ${\bf{J}}_{constitutive}={\bf{J}}_{Ohm}$. Its temporal Fourier transformation was known as early as 1928 by communication engineers like John Renshaw Carson from Bell System \cite{Carson}. Indeed, from the temporal Fourier transformation of the  Maxwell-Amp\`{e}re equation $\nabla \times {\bf{\hat{H}}}=i\omega \epsilon {\bf{\hat{E}}}+\sigma {\bf{\hat{E}}}$, Carson introduced a complex permittivity $\underline{\epsilon}= \epsilon -i\frac{\sigma}{\omega}$ into the temporal Fourier transformation of the Lorenz "gauge condition" $\nabla .{\bf{A}}+\mu \epsilon \frac{\partial V}{\partial t} =0$ \cite{Okun, Lorenz} to obtain the temporal Fourier transformation of the Stratton "gauge condition" $\nabla .{\bf{\hat{A}}}+(i\omega \mu \epsilon +\mu \sigma )\hat{V}=0$.

According to Stratton's alternative procedure, Gauss' law $\nabla .{\bf{E}}=0$ implies immediately:
\begin{equation}
\nabla ^2V + \frac{\partial }{\partial t} \left(\nabla .{\bf{A}}\right)=0
\end{equation}
which can be solved if and only if the potentials are constrained by the Stratton "gauge condition":
\begin{equation}
\nabla .{\bf{A}}+\mu \epsilon \frac{\partial V}{\partial t} =-\mu  \sigma V.
\end{equation}
In the simple case of constant permeability $\mu$ and permittivity $\epsilon$, Stratton deduced from the Maxwell-Minkowski's set the following equations (${\bf S}_{Stratton}=0$):
\begin{equation}
\nabla ^2V-\mu \epsilon \frac{\partial ^2V}{\partial t^2} -\mu  \sigma  \frac{\partial V}{\partial t}=0
\end{equation}
\begin{equation}
\nabla ^2{\bf{A}}-\mu \epsilon \frac{\partial ^2{\bf{A}}}{\partial t^2} -\mu  \sigma  \frac{\partial {\bf{A}}}{\partial t}=-\mu {\bf{J}}_{external}
\end{equation}
which are the well-known "telegrapher's equations". They were derived previously for the tension and the current by Vaschy and Heaviside starting from the global electrical equations of Kirchhoff for circuitry and not directly from the local Maxwell-Minkowski equations for the fields. As an example, they described the propagation of waves in a coaxial cable with Ohmic dissipation. Later, Paul Poincelot derived its tensorial expression since the Stratton "gauge condition" is not manifestly Relativistic covariant under the Lorentz transformations of space-time \cite{Poincelot}. The more famous Lorenz "gauge condition" \cite{Okun, Lorenz} is the dissipation-free version of the Stratton's constraint ($\sigma =0$). As a partial conclusion, it is very surprising to notice that the Stratton "gauge condition" is completely absent from modern textbooks and is not even mentioned in the benchmark review paper on the history of Gauge Invariance \cite{Okun}.

Now, we recall the reader of the physical meaning of the potentials \cite{A} and their constraints. As for the Stratton "gauge condition", the following interpretations of the "gauge conditions" are nowhere in modern treatments of Classical Electromagnetism. The Lorenz "gauge condition" for vacuum $\nabla .{\bf{A}}+\frac{1}{c^2} \frac{\partial V}{\partial t} =0$ is analogous to the mass continuity equation for compressible flows in the particular case of the linearized acoustic perturbations. As a matter of fact, the mass conservation of a flowing fluid is encoded in the following law \cite{GHP}:
\begin{equation}
\nabla .\left(\rho {\bf{u}}\right)+ \frac{\partial \rho}{\partial t} =0.
\end{equation}
If we perturb the density, pressure and velocity around a basic state at rest:
$\rho = \rho _0 +\delta \rho$, $p=p_0+\delta p$ and ${\bf{u}}={\bf{0}}+\delta {\bf{u}}$, the continuity equation can be recast in a Lorenz "gauge condition" form:
\begin{equation}
\nabla .\left(\delta {\bf{u}}\right)+ \frac{1}{c_s^2}\frac{\partial }{\partial t}\left(\frac{\delta p}{\rho _0}\right) =0
\end{equation}
where $c_s=\frac{1}{\sqrt{\rho \kappa}}=\sqrt{\frac{\partial p}{\partial \rho}}\simeq \sqrt{\frac{\delta p}{\delta \rho}}$ is the speed of sound analogous to the speed of light in vacuum $c=\frac{1}{\sqrt{\mu _0 \epsilon _0}}$.

The Coulomb "gauge condition" $\nabla .{\bf{A}}=0$ is analogous to the mass continuity equation for incompressible flows $\nabla .{\bf{u}}=0$ \cite{GHP} provided that the compressibility (permittivity) vanishes i.e. $\kappa \to 0$ at constant density $\rho _0$ (permeability). As we will see later on, this approximation corresponds to the Galilean (magnetic) limit of the Lorenz "gauge condition"  \cite{Montigny, EPL05, AJP}.

The Stratton "gauge condition" is a generalized continuity equation for the vector potential :
\begin{equation}
\nabla .{\bf{A}}+\mu \epsilon \frac{\partial V}{\partial t} =-\mu  \sigma V.
\end{equation}
The right-hand side is a sink term. The vector potential is dissipated by Ohmic conduction. Loci of high scalar potential are sinks for the vector potential whose flux is directed towards them. The Stratton "gauge condition" is analogous to the mass continuity equation with nuclear reactions acting as a sink.

Thanks to the above analogy with Fluid Mechanics, it is now obvious to the reader that the vector (scalar) potential is a kind of electromagnetic momentum (energy) per unit charge \cite{A}. Once again, modern Physics has almost completely forgotten the physical meaning of the potentials as it was formulated by James Clerk Maxwell in the nineteenth century and part of his results are rediscovered from time to time either by historians of science or Physics teachers \cite{A}.

We have just recalled three examples of "gauge conditions". It is clear that the analogy with Fluid Mechanics advocates for different domains of validity depending on the underlying Physics. Here, we will discuss how to choose a "gauge condition" depending on the context. Our method will be dimensional analysis as often in Fluid Mechanics. Our guide will be Relativistic or Galilean Covariance. That is why we start by a recap on Galilean Electromagnetism as described by Physicists following L\'evy-Leblond and Le Bellac \cite{LBLL, Montigny, EPL05, AJP, EPL08} and Engineers following another M.I.T. researcher James Melcher \cite{Melcher}.

We list first the dimensional quantities. An electromagnetic phenomenon happens in a spatial arena of extension $L$ in a duration $\tau$. The arena is a continuous medium with constitutive properties $\epsilon, \mu$ and $\sigma$ taken as constant for simplicity (otherwise they are tensors with time and space dependance). Applying the Vaschy-Buckingham theorem of dimensional analysis \cite{GHP}, we can construct dimensionless parameters which would characterize the electromagnetic response of the continuous medium. As we will deal with Galilean approximations, we introduce $v\approx \frac{L}{\tau}$ the typical velocity of the system and we compare it with $c_m=\frac{1}{\sqrt{\mu \epsilon}}$ the light celerity in the continuous medium. The Galilean limit (quasi-static approximation) corresponds to $v<<c_m$. If we neglect time dependance in the Stratton system ${\bf S}_{Stratton}=0$ ($\partial /\partial t =0$ or $\partial /\partial t \simeq 0$), we get $\nabla ^2{\bf{A}}=-\mu {\bf{J}} \quad and \quad \nabla ^2V=-\frac{\rho}{\epsilon}$. In terms of orders of magnitude \cite{LBLL, Montigny, EPL05, AJP, EPL08} (the tilde means order of magnitude), we deduce $\tilde{A} \approx L^2 \mu \tilde{J} \quad and \quad \tilde{V} \approx \frac{L^2}{\epsilon}\tilde{\rho}$. Hence, we construct by hand the dimensionless parameter:
\begin{equation}
\frac{ \frac{1}{\sqrt{\epsilon \mu }} \tilde{A} }{ \tilde{V} } \approx \frac{ \tilde{J}}{ \tilde{\rho} \frac{1}{\sqrt{\epsilon \mu }} }
\end{equation}
which characterizes the type of regime \cite{LBLL, Montigny, EPL05, AJP, EPL08}: (i) $v\simeq c_m$ and $c_m \tilde{A} \simeq \tilde{V}$ $\to$ Relativistic regime;  (ii)  $v<< c_m$ and  $c_m \tilde{A} >>\tilde{V}$ ($\tilde{V}\simeq v\tilde{A}$) $\to$ Galilean magnetic limit (magnetoquasi-statics or MQS); (iii)  $v<< c_m$ and  $c_m \tilde{A} <<\tilde{V}$ ($\tilde{A}\simeq \frac{v}{c_m^2}\tilde{V}$) $\to$ Galilean electric limit (electroquasi-statics or EQS).

The Stratton's continuity equation becomes (the bar denotes a dimensionless quantity):
\begin{equation}
\frac{\tilde{A}}{L} \underline{\nabla} .{\bf{ \underline{A}}} + \mu \epsilon \frac{\tilde{V}}{\tau}  \frac{\partial \underline{V}}{\partial \underline{t}}=- \mu  \sigma \tilde{V}\underline{V}
\end{equation}
that is:
\begin{equation}
\underline{\nabla} .{\bf{ \underline{A}}} + \frac{\frac{L}{\tau}}{\frac{1}{\sqrt{\mu \epsilon}}} \frac{ \tilde{V} }{ \frac{1}{\sqrt{\epsilon \mu }} \tilde{A} } \frac{\partial \underline{V}}{\partial \underline{t}}=- \frac{L}{l_*} \frac{ \tilde{V} }{ \frac{1}{\sqrt{\epsilon \mu }} \tilde{A} }\underline{V}
\end{equation}
whose mathematical form is simply $I+II=III$ with the following dimensionless ratios:
\begin{equation}
\frac{II}{I}\approx \frac{v}{c_m} \frac{ \tilde{V} }{c_m \tilde{A}} =\frac{\tau _{em}}{\tau} \frac{ \tilde{V} }{c_m \tilde{A}}
\end{equation}
\begin{equation}
\frac{III}{I}\approx \frac{L}{l_*} \frac{ \tilde{V} }{ c_m \tilde{A}} =\frac{\tau _m}{\tau _{em}} \frac{ \tilde{V} }{c_m \tilde{A}}=\frac{\tau _{em}}{\tau _{e}} \frac{ \tilde{V} }{c_m \tilde{A}}
\end{equation}
\begin{equation}
\frac{III}{II}\approx \frac{\tau}{\tau _e}
\end{equation}
and where we introduced the following parameters \cite{Melcher}: $l_*=\frac{1}{\sigma}\sqrt{\frac{\epsilon}{\mu}}$ the constitutive length, $\tau _{em}= \frac{L}{c_m}$ the light transit time, $\tau _e= \sqrt{\frac{\epsilon}{\sigma}}$ the charge relaxation time and $\tau _m= \mu \sigma L^2$ the magnetic diffusion time such that $\tau_{em}= \sqrt{\tau_e \tau _m}$. 

\begin{figure}[!htbp]
\includegraphics[width=8cm]{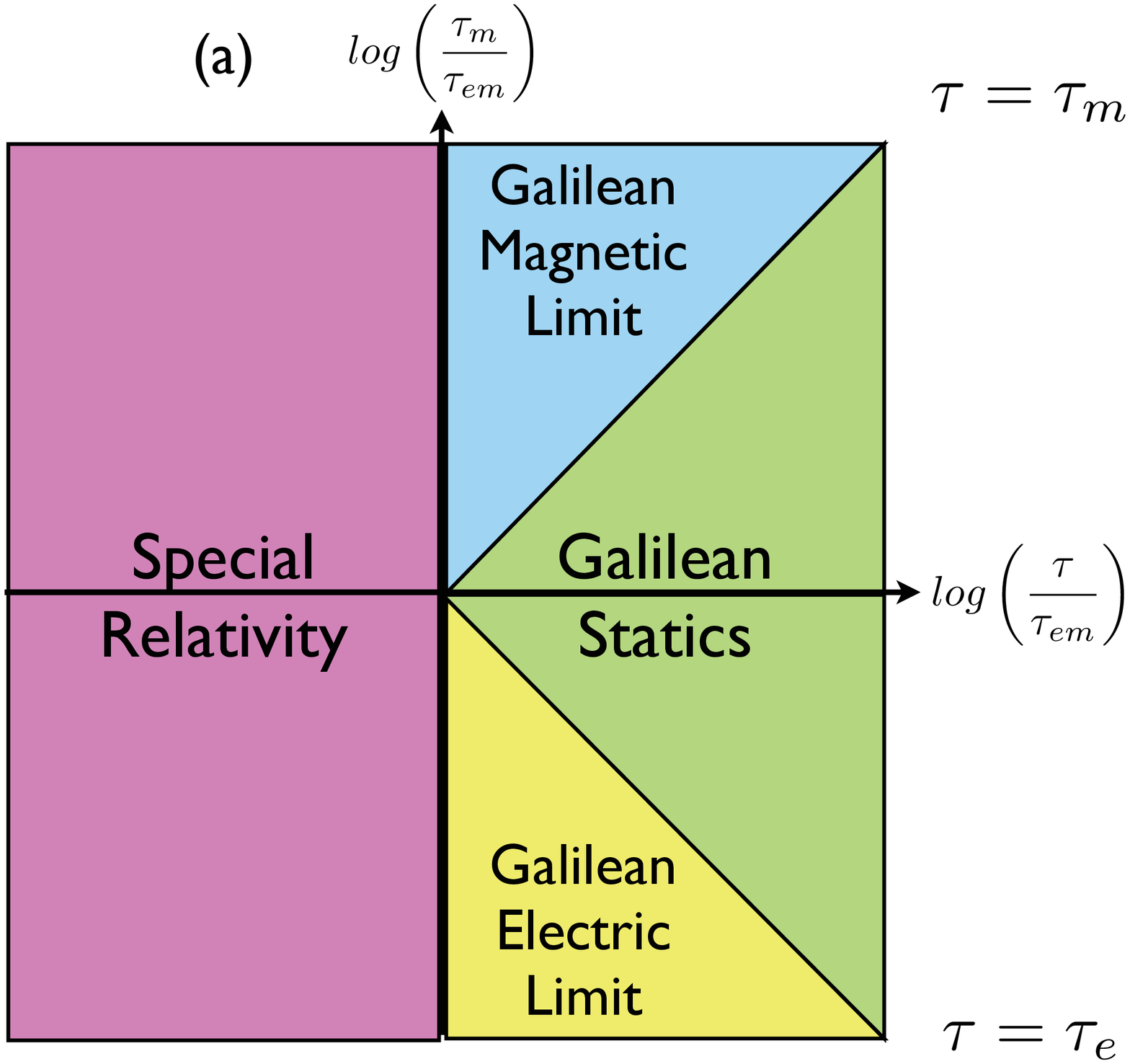}
\includegraphics[width=8cm]{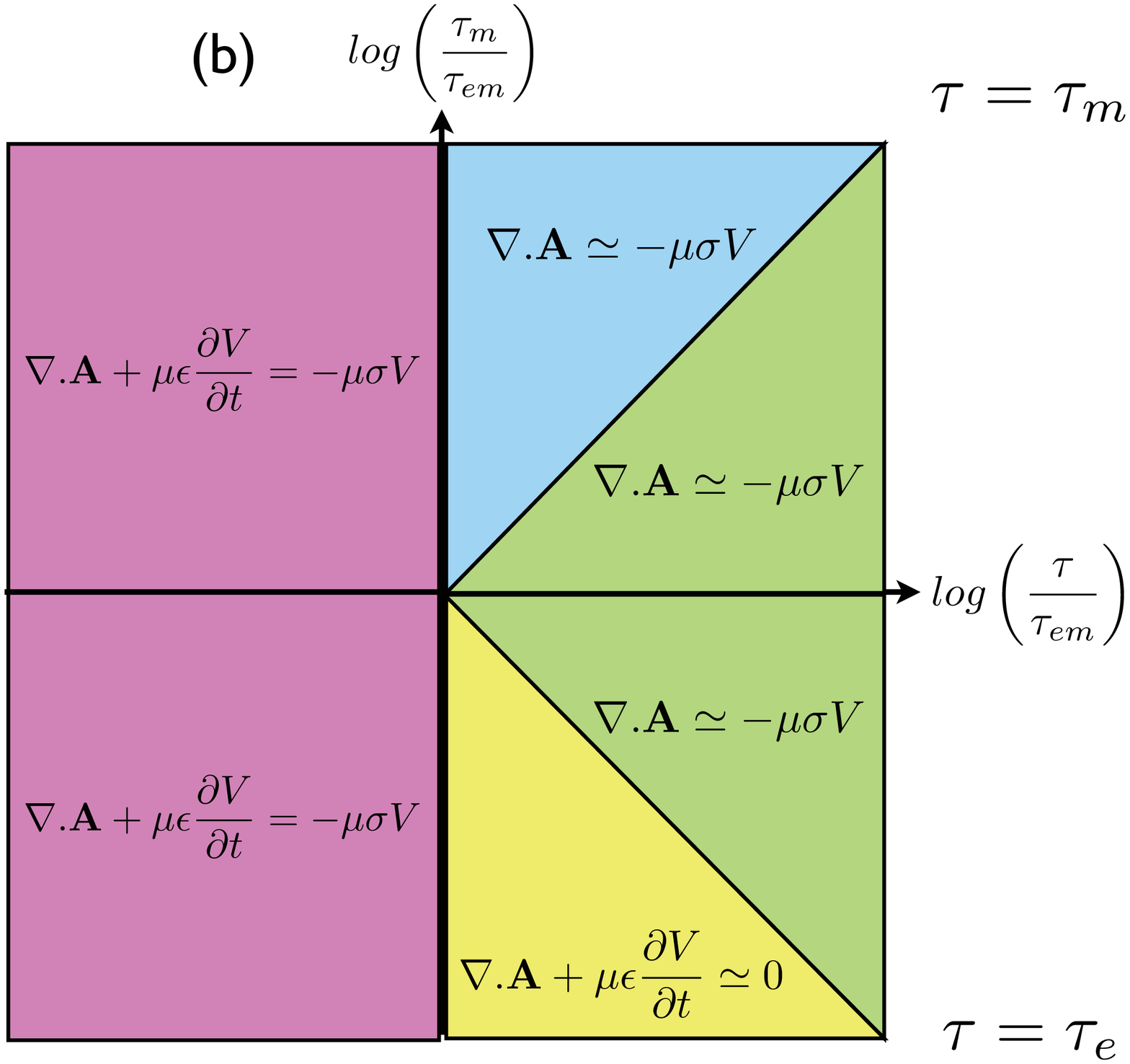}
\caption{Domains of validity of (a) the Galilean limits and (b) the various "gauge conditions" in a log-log plot inspired by Melcher \cite{Melcher} with dimensionless times as variables.}
\label{gauges}
\end{figure}

The Figure \ref{gauges} displays the different approximations of the Stratton's constraint depending on the Relativistic or Galilean (Magnetic, Electric or Statics) regime for a given problem. In practice, we compare the magnitude of the three terms $I$, $II$ and $III$ in the Stratton's constraint using the scaling laws (i), (ii) or (iii).

Hence, the "gauge conditions" are continuity equations whose domains of validity depend on the Relativistic or Galilean nature of the underlying phenomenon and have nothing to do with mathematical closure assumptions taken without physical motivations.

According to our results, Gauge Invariance is NOT a fundamental symmetry of Physics since (1) the "gauge transformations" can be avoided by a direct definition of the potentials as mathematical solutions of the Maxwell-Minkowski equations; (2) the "gauge conditions" are interpreted physically as electromagnetic continuity equations; (3) the "gauge fields" are interpreted physically as electromagnetic energy and momentum per unit charge; (4) the "gauge conditions" have domains of validity derived from Relativistic or Galilean Covariance.\\

The author would like to thank Francesca Rapetti for playing the role of a sounding board.

\end{document}